\documentclass[pre,showpacs,preprintnumbers,amsmath,amssymb,superscriptaddress]{revtex4-1}

\pdfoutput=1

\usepackage{graphicx}
\usepackage{hyperref}
\usepackage{bm}

\usepackage{amsmath}
\usepackage{amssymb}
\usepackage{latexsym}
\usepackage{multirow}

\usepackage{color}
\definecolor{dockerblue}{rgb}{0.11,0.56,0.98}
\definecolor{oneblue}{rgb}{0,0,0.75}

\begin{document}

\title{Dispersion in two dimensional channels - the Fick-Jacobs approximation revisited}
\author{M. Mangeat}
\author{T. Gu\'erin}
\author{D. S. Dean}
\affiliation{Laboratoire Ondes et
Mati\`ere d'Aquitaine (LOMA), CNRS, UMR 5798 / Universit\'e de Bordeaux, F-33400 Talence, France}

\bibliographystyle{apsrev}

\begin{abstract}
We examine the dispersion of Brownian particles in a symmetric two dimensional channel, this classical problem has been widely studied in the literature using the so called Fick-Jacobs' 
approximation and its various improvements. Most studies rely on the reduction to an effective one dimensional diffusion equation, here we derive an explicit formula for the diffusion constant 
which avoids this reduction. Using this formula the effective diffusion constant can be evaluated
numerically without resorting to Brownian simulations. In addition, a perturbation theory can be
developed in $\varepsilon = h_0/L$ where $h_0$ is the characteristic channel height and $L$ the period. This perturbation theory confirms the results of Kalinay and Percus (Phys. Rev. E {\bf 74}, 041203 (2006)), based on the reduction, to one dimensional diffusion are exact at least to  ${\cal O}(\varepsilon^6)$. Furthermore, we show how the Kalinay and Percus pseudo-linear approximation can be straightforwardly recovered. The approach proposed here can also be exploited to yield exact results in the limit $\varepsilon \to \infty$, we show that here the diffusion constant remains finite and show how the result can be obtained with a simple physical argument. Moreover, we show that the 
correction to the effective diffusion constant is of order $1/\varepsilon$ and remarkably has a some universal characteristics. Numerically we compare the analytic results obtained with exact numerical calculations for a number of interesting channel geometries.

\end{abstract}

\maketitle
\section{Introduction}
The dispersion properties of complex systems have a wide range of applications including solid state physics, soft matter physics, biology and chemical engineering \cite{van2007,gar1985}. 
In many cases the effective transport properties, such as the late time diffusion constant and effective drift are important quantities to study problems as varied as 
 mixing \cite{leBorgne2013stretching,dentz2011mixing,barros2012flow}, sorting \cite{bernate2012stochastic}, contaminant spreading \cite{brusseau1994transport} or reaction kinetics in complex systems \cite{con2007}. The computation of such effective transport quantities is a longstanding problem in statistical physics and exploits techniques ranging from probability theory \cite{kel1962} through to quantum field theory \cite{dea2007}. 

Transport, notably diffusion, in channel like geometries arises in a number of important physical contexts, ranging from biophysics through to the study of porous media \cite{burada2009diffusion,malgaretti2013entropic,bressloff2013stochastic,holcman2013control}. Dispersion in periodic systems such as hard squares or disks placed on a lattice can also be mapped onto diffusion in channel like systems \cite{dag2012b}.
Even if the underlying process in such problems is locally standard Brownian motion, the interplay between geometry and diffusion gives rise to rich behavior, for instance the slowing down of diffusive dispersion due to entropic trapping. In many contexts one is only interested in dispersion along the channel and the problem of computing the effective diffusion constant parallel to the channel is one which has been intensively studied. The
first results on diffusion in channels are attributed to Jacobs in Ref. \cite{jac1967} where the first form of what is known as the Fick-Jacobs' (FJ) approximation was derived. Subsequent works \cite{zwa1991,reg2001,kal2005,kal2006,mar2011,bra2009,ber2011a,dag2012} involved improvements to the basic FJ approximation for both two dimensional channels and axisymmetric channels in three dimensions. These studies are based on the effective reduction of the problem of diffusion in the channel, to a one dimensional problem where a particle diffuses in an entropically generated potential, simply related to the local channel width or area, but with an effective local diffusion constant. Once this reduction is carried out the effective diffusion constant can be computed using exact results for the diffusion constant
of one-dimensional systems \cite{lif1961,rei2001,costantini1999}. 

One of the most successful approaches to the problem of diffusion in a channel was proposed
by Kalinay and Percus in Refs.~\cite{kal2005,kal2006}. 
Here they express the effective local diffusion constant as an expansion in powers of a parameter $\varepsilon$ proportional to the ratio of the typical relaxation times in the perpendicular and parallel direction. Later studies \cite{dorfman2014assessing} checked that such expansions lead to expressions of the global effective diffusivity that are exact at least to order $\mathcal{O}(\varepsilon^4)$. 
From a probabilistic point of view this latter result is somewhat surprising as the effective 
process parallel to the channel can clearly never be Markovian and described by an effective one dimensional diffusion equation. Mathematically similar problems, where a higher dimensional Markovian stochastic processes is reduced to a lower dimensional one, occur throughout the physics literature, notably when partially-damped Langevin equations 
(describing the velocity and position of a particle) are reduced to over-damped Langevin equations (describing only the position). A similar problem arises for a particle carrying a dipole in a spatially varying electric field, here the Langevin equations describing the dipole moment and the position can be  reduced to a Langevin equation for just the position.  In the class of problems mentioned above, a mathematical process of adiabatic elimination, which amounts to a perturbation expansion in the time scale for the relaxation of the quick variable (velocity or dipole moment for instance) divided by that of the slow position variable can be applied. In most cases the effective process remains Markovian \textit{only} to first order in perturbation theory \cite{mig1980,ris1996,tou2009}. 
In this paper we neglect hydrodynamic effects, these can become important for fluid filled channels \cite{yang2017} when the particle size becomes comparable to the channel size. This is due to the spatial variation of the local diffusion tensor as one approaches the walls channel. Here we  restrict our attention to   the dispersion of a point-like  particle and we can thus ignore these hydrodynamic effects.


In this paper, we analyze in section \ref{Sec1} the FJ problem starting from an exact formula for the longitudinal diffusion constant. The formula is derived here using a physical argument for channels in any dimension, and agrees with general Kubo type formulas derived for advection diffusion equations in periodic systems \cite{gue2015a,gue2015b} and with the macroscopic transport theory developed by Brenner \cite{bre1993}. Then, for two-dimensional channels, we present in section \ref{Sec2} a very compact analysis which allows us to show in section \ref{Sec3} that the results of Ref.~\cite{kal2006} are exact to ${\cal O}(\varepsilon^6)$, whereas to date they had been checked only up to ${\cal O}(\varepsilon^4)$ by using Brenner's theory \cite{dorfman2014assessing}. Given what we have stated about adiabatic approximations above, it is very interesting that the method of Ref.~\cite{kal2006} stays correct up to such a high order.  We obtain this result within a formalism that uses analytic functions in a complex plane, and we show in section \ref{Sec4} how the pseudo-linear approximation of Kalinay and Percus can be recovered. This approximation is based on neglecting terms $W''(x)$ (where $W$ is the local channel width) and higher order derivatives. In section \ref{Sec5}, we show how the diffusion constant can be computed exactly for a class of two dimensional channels which can be appropriately parameterized in the complex plane, this reproduces in a very rapid manner the classical result of Zwanzig \cite{zwa1982}. We also show in section \ref{Sec6} how the same formalism can be applied to the  limit $\varepsilon \to \infty$, the exact opposite of the cases usually studied in the literature and in which all approaches based on a dimensional reduction fail. In this limit $\varepsilon\to\infty$, we find the asymptotic value of the diffusion constant (which is non-zero) and we also compute the leading order correction, which is ${\cal O}(1/\varepsilon)$, and which has a prefactor that has a remarkably universal character. A short description of this result was given in \cite{man2017}, the derivation presented here is slightly different, as we concentrate here on two dimensional channels, and relies more heavily on complex analysis. Finally, in section \ref{Sec7} we show how the formalism based on the Kubo formula for the diffusion constant can be used to numerically study a wide range of channel geometries, thus avoiding the statistical errors and longer times associated with Brownian simulations. We compare some of the standard variants of the FJ approximation with these results for a number of channel geometries, as well as the high order perturbation theory and large $\varepsilon$ results derived here. 

%




\section{General equations for diffusion in channels}
\label{Sec1}

\subsection{Model}
In this section, we consider diffusion in a portion of space located between two surfaces of equations $y=h_1({\bf x})$ and $y=h_2({\bf x})$. Here ${\bf x}$ represents the  coordinates in the hyperplane parallel to the channel, and $y$ is the coordinate in the perpendicular direction. 
We consider the diffusion of a tracer with isotropic diffusion in the channel, with diffusion constant $D$, and subject to a uniform drift velocity parallel to the channel ${\bf u}_{||}$. The 
probability density function (pdf) for the tracer thus obeys the advection-diffusion equation
\begin{equation}
\frac{\partial p}{\partial t}({\bf x},y;t) = \nabla\cdot \left[ D \nabla p({\bf x},y;t)- {\bf u}_{||} p({\bf x},y;t)\right].\label{deq}
\end{equation}
with no-flux boundary condition at the interface. 
We now define the marginal pdf for the coordinate ${\bf x}$ as
\begin{equation}
p^*({\bf x};t) = \int_{h_1({\bf x})}^{h_2({\bf x})} dy\ p({\bf x},y;t)\label{pstar}
\end{equation}
The basis of FJ type approximations is to find an effective diffusion equation for the marginal pdf in the case where it describes a one dimensional process. In what follows, we will (i) derive an exact equation for $p^*({\bf x},t)$ to make the hypotheses of the FJ approximation more explicit, and (ii) we will derive an exact equation for the effective diffusivity. 

\subsection{An exact equation for the evolution of the marginal distribution}

 Here, we shall derive an exact equation for $p^*$, which makes more explicit the approximations of the FJ approach. To proceed we take the time derivative of Eq.~\eqref{pstar} and use the diffusion equation
\eqref{deq} to obtain 
\begin{eqnarray}
\frac{\partial p^*}{\partial t}({\bf x};t) &=& \int_{h_2({\bf x})}^{h_1({\bf x})} dy \left[\nabla_{||}\cdot (D\nabla_{||} p({\bf x},y;t) -{\bf u}_{||} p({\bf x},y;t))
+ D\frac{\partial^2}{\partial y^2}p({\bf x},y;t) \right]\nonumber \\
&=& \int_{h_2({\bf x})}^{h_1({\bf x})} dy \nabla_{||}\cdot (D\nabla_{||} p({\bf x},y;t) -{\bf u}_{||} p({\bf x},y;t)) + D\frac{\partial p}{\partial y}({\bf x},h_2({\bf x});t) - D \frac{\partial p}{\partial y}({\bf x},h_1({\bf x});t) . 
\end{eqnarray}


The no-flux boundary condition at each surface is given explicitly as at the surface $y=h_{\alpha}({\bf x})$ (for $\alpha = 1,\ 2$):
\begin{equation}
D\frac{\partial p}{\partial y}({\bf x},h_\alpha({\bf x});t) = [D \nabla_{||}p({\bf x},y;t) -{\bf u}_{||} p({\bf x},y;t)]|_{y=h_\alpha({\bf x})} \cdot \nabla_{||}h_\alpha({\bf x}),\label{bc}
\end{equation}
where we have used the decomposition of the gradient operator as $\nabla = \nabla_{||} + {\bf e}_y \frac{\partial}{\partial y}$. This yields
\begin{eqnarray}
\frac{\partial p^*}{\partial t}({\bf x};t) 
&=& \int_{h_2({\bf x})}^{h_1({\bf x})} dy \nabla_{||}\cdot [D\nabla_{||} p({\bf x},y;t) -{\bf u}_{||} p({\bf x},y;t)] \nonumber \\ &+&[D \nabla_{||}p({\bf x},y;t) -{\bf u}_{||} p({\bf x},y;t)]|_{y=h_2({\bf x})}\cdot \nabla_{||}h_2({\bf x}) - [D \nabla_{||}p({\bf x},y;t) -{\bf u}_{||} p({\bf x},y;t)]|_{y=h_1({\bf x})} \cdot \nabla_{||}h_1({\bf x}) .\label{step}
\end{eqnarray}
Now using the fact that ${\bf u}_{||}$ is uniform we can write
\begin{eqnarray}
\nabla_{||} \cdot [ {\bf u}_{||} p^*({\bf x};t)]=\nabla_{||} \cdot \int_{h_1({\bf x})}^{h_2({\bf x})} dy \ {\bf u}_{||} p ({\bf x},y;t)
&=& \nabla_{||} h_2({\bf x})\cdot {\bf u}_{||} p({\bf x},h_2({\bf x});t) - \nabla_{||} h_1({\bf x})\cdot {\bf u}_{||} p({\bf x},h_1({\bf x});t) \nonumber\\&+& \int_{h_1({\bf x})}^{h_2({\bf x})} dy \nabla_{||} \cdot[{\bf u}_{||} p({\bf x},y;t)],
\end{eqnarray}
and we also find
\begin{eqnarray}
\nabla_{||}^2 p^*({\bf x};t)&=&\nabla_{||}^2 \int_{h_1({\bf x})}^{h_2({\bf x})} dy \ p({\bf x},y;t) =  \nabla_{||}\cdot [ p({\bf x},h_2({\bf x});t) \nabla_{||}h_2({\bf x})]+ \nabla_{||}h_2({\bf x})\cdot\nabla_{||}p({\bf x},y;t))|_{y=h_2({\bf x})} \nonumber \\&-&\nabla_{||}\cdot [ p({\bf x},h_1({\bf x});t) \nabla_{||}h_1({\bf x})]-\nabla_{||}h_1({\bf x}) \cdot\nabla_{||}p({\bf x},y;t))|_{y=h_1({\bf x})} + \int_{h_2({\bf x})}^{h({\bf x})} dy\ \nabla_{||} ^2 p({\bf x},y;t).
\end{eqnarray}
Using these results Eq.~\eqref{step} simplifies remarkably to give
\begin{equation}
\frac{\partial p^*}{\partial t}({\bf x};t) = D\nabla_{||}^2 p^*({\bf x};t) - \nabla_{||}\cdot [{\bf u}_{||}
p^*({\bf x};t)] - D\nabla_{||}\cdot \left[p({\bf x}, h_2({\bf x});t) \nabla_{||} h_2({\bf x})-p({\bf x}, h_1({\bf x});t) \nabla_{||} h_1({\bf x})\right].\label{exact}
\end{equation}
The above equation is exact and is the main result of this section. Even if it is not a closed equation for $p^*$, we see  that the evolution of the marginal pdf $p^*$ involves only the value of the full probability density at the surfaces $y= h_\alpha({\bf x})$, not in the whole channel. 

\subsection{The basic Fick-Jacobs' approximation}
The basic FJ approximation is obtained by assuming that $p({\bf x},y,t) \simeq p_0({\bf x},t)$, {\em i.e.} assuming that it is independent of $y$, and thus one is assuming that the pdf equilibrates in the coordinate $y$. This means that
\begin{equation}
p^*({\bf x};t) \simeq \int_{h_1({\bf x})}^{ h_2({\bf x})} dy\ p_0({\bf x};t) = H({\bf x})p_0({\bf x};t) ,
\end{equation}
where $H({\bf x}) = h_2({\bf x}) -h_1({\bf x})$ is the total channel width at the in-plane coordinate ${\bf x}$. The basic FJ approximation then gives
\begin{equation}
p({\bf x},h_1({\bf x});t) \simeq p({\bf x},h_2({\bf x});t) \simeq \frac{p^*({\bf x};t)}{H({\bf x})}
\end{equation}
and so the effective diffusion equation for $p^*({\bf x};t)$ becomes
\begin{equation}
\frac{\partial p^*}{\partial t}({\bf x};t) = D\nabla_{||}^2 p^*({\bf x};t) - \nabla_{||}\cdot {\bf u}_{||}
p^*({\bf x};t) - D\nabla_{||}\cdot \left(p^*({\bf x};t) \nabla_{||} \ln[H({\bf x}]\right).\label{fj0}
\end{equation}
The physical validity of this approximation is argued to hold when the process relaxes in the 
$y$ direction very quickly with respect to its motion in the direction parallel to the channel; it
should hold when $H({\bf x}) \ll L$ where $L$ is the length scale over which the total channel
width varies. An alternative limit where it holds is when one takes the diffusion to be anisotropic, having a (isotropic) diffusion constant $D_{||}$ in the plane and a different diffusion constant $D_\perp$ in the perpendicular direction. The FJ approximation then holds in the limit $\varepsilon =D_{||}/D_\perp\ll 1$, and the problem can be analyzed perturbatively using $\varepsilon$ as the small parameter. Of course the approximation that $p({\bf x},y;t)$ is independent of $y$ clearly does not satisfy the no flux boundary condition in Eq.~\eqref{bc} when $\nabla_{||}h_\alpha\neq {\bf 0}$. An important success of the FJ approximation is that it gives the correct equilibrium distribution in the steady state. On a unit cell $\Omega$ of the channel (which is repeated to form the infinite channel) the equilibrium distribution is clearly uniform within the channel and thus the marginal pdf for the in plane coordinate ${\bf x}$ is 
\begin{equation}
P_s({\bf x}) = \frac{H({\bf x})}{\int_\Omega d{\bf x'} H({\bf x'})}.\label{ssp}
\end{equation}
This means that the ${\bf x}$ in-plane coordinate of the tracer particle is subjected to an effective entropic potential
\begin{equation}
\phi_{\rm ent}({\bf x}) = -k_BT\ln(H({\bf x})).
\end{equation}

For effectively two dimensional systems (that is to say two dimensional channels or three 
dimensional ones with rotational symmetry) a number of authors have suggested improvements to the Fick-Jacobs approximation in the form an effective in plane diffusion constant $D(x)$ which depends on the local height functions (in general both $h_2$ and $h_1$ and not just their difference). Clearly if an underlying one dimensional diffusion equation were to exist it would have to have the steady state distribution given by Eq.~(\ref{ssp}) and thus have the general form, in the absence of external drift,
\begin{equation}
\frac{\partial p^*}{\partial t}(x;t) = \frac{\partial}{\partial x} D(x) \left[ \frac{\partial p^*}{\partial x}(x;t)
+\beta \frac{\partial \phi_{\rm ent}}{\partial x}(x) p^*(x;t)\right], 
\end{equation}
with $\beta=1/k_BT$. The effective diffusion constant for the effective one dimensional problem can then be extracted using the exact formula for one dimensional diffusion \cite{lif1961}
\begin{equation}
D_e = \frac{1}{\langle D^{-1}(x)\exp\left(\beta\phi_{\rm ent}(x)\right)\rangle_s \langle \exp\left(-\beta\phi_{\rm ent}(x)\right)\rangle_s},\label{lif}
\end{equation}
where the angled bracket indicate the spatial average over one period of the system, {\em i.e.}
\begin{equation}
\langle f\rangle_s = \frac{1}{L}\int_0^L dx \ f(x).
\end{equation}
These approaches all agree in an expansion based on $\varepsilon= D_\perp/D_{||}$ being small but differ at higher orders. However, as mentioned in the introduction, this approach clearly cannot be valid exactly as the process projected onto the in-plane coordinates is evidently not Markovian and thus cannot be described by a diffusion equation with local transport coefficients. As mentioned in the introduction, a similar phenomenon appears in the theory of stochastic processes where a fast variable is eliminated from a stochastic differential equation using an adiabatic approximation scheme \cite{mig1980,ris1996,tou2009}. It is generally found that the process remains Markovian at the first order of correction but does not remain so when higher order corrections are included. 

\subsection{General formula for the diffusion constant}
In this paper we will use an exact formula for the diffusion constant, of the process in the plane of the channel, which
does not use a reduction of the problem to a lower dimensional one. The formulas for the diffusion constant can be 
directly written down using a general theory for diffusion in periodic systems developed in \cite{gue2015a,gue2015b}.
These formulas are formally exact -- however, they require the solution of partial differential equations which are
not analytically soluble in two dimensions and higher. Nevertheless, the relevant equations can be treated using approximation methods for narrow channel geometries which are similar to those which are employed to derive improvements to the 
FJ approximation and which are used as the basis for lubrication theory in fluid dynamics \cite{lub}. 

From now on, we consider that $h_1$ and $h_2$ are spatially periodic, and of the same periodicity. Rather than simply quoting the results of \cite{gue2015a,gue2015b}, we provide a quick re derivation of the result, ideally suited to the channel geometry and based on simple physical arguments. To begin with we compete the average drift of 
a tracer in the in plane coordinates,
\begin{equation}
\frac{d\langle X_i\rangle}{dt} = \int d{\bf x} dy \ x_i \frac{\partial p}{\partial t}({\bf x},y;t) = \int d{\bf x}\ x_i \frac{\partial p^*}{\partial t}({\bf x},t),
\end{equation}
and we now use Eq.~({\ref{exact}) and integration by parts to give an effective drift
\begin{equation}
\frac{d\langle {\bf X}\rangle}{dt} = {\bf u}_{||} + D \int d{\bf x} \left[ p\left({\bf x},h_2({\bf x});t\right) \nabla_{||}h_2({\bf x}) -
p\left({\bf x},h_1({\bf x});t\right) \nabla_{||}h_1({\bf x}) \right].
\end{equation}
In the steady state we can replace $p\left({\bf x},h_2({\bf x});t\right)$ by their steady state values on the unit cell $\Omega$,
denoted by $P_s({\bf x},y)$ and use periodicity to restrict the ${\bf x}$ integral over the in-plane coordinates of $\Omega$ denoted by $\Omega_{||}$, to find
\begin{equation}
\frac{d\langle {\bf X}\rangle}{dt} = {\bf u}_{||} + D \int_{\Omega_{||}} d{\bf x} \left[ P_s\left({\bf x},h_2({\bf x})\right) \nabla_{||}h_2({\bf x}) -
P_s\left({\bf x},h_1({\bf x})\right) \nabla_{||}h_1({\bf x}) \right].
\end{equation}
Now we  suppose that  the drift ${\bf u}_{||}$ is induced by an infinitesimal applied force ${\bf F}_{||}$, i.e. ${\bf u}_{||} = D\beta {\bf F}_{||}$. This local constant applied force will give rise to an effective late time, or asymptotic, drift which is defined by $V_{||i} = d \langle X_i \rangle/dt$.  The Stokes Einstein relation then gives
\begin{equation}
V_{||i} = \beta D_{e}^{ij}{F}_{||j} = \frac{D_{e}^{ij}}{D}{u}_{||j},
\end{equation}
where the Einstein convention on index summation is used, and $D_{e}^{ij}$ is the effective diffusion tensor in the in-plane coordinates, given for ${\bf u}_{||}=0$ by
\begin{align}
D_{e}^{ij}=\lim_{t\rightarrow\infty} \frac{\langle x_i(t)x_j(t)\rangle}{2t}
\end{align}
The steady state distribution obeys
\begin{equation}
\nabla\cdot \left[ D \nabla P_s({\bf x},y)- {\bf u}_{||} P_s({\bf x},y)\right]=0, \label{ss}
\end{equation}
and then assuming that $u_\perp=0$ and ${\bf u}_{||}$ is small we can write
\begin{equation}
P_s({\bf x},y) = \frac{1}{\Omega} + u_{||j}\frac{\partial P_s}{\partial u_{||j}}({\bf x},y),
\end{equation}
where we have also used $\Omega$ to denote the volume of the unit cell $\Omega$.
We write $f_i = D\frac{\partial P_s}{\partial u_{||i}}$ and so the Stokes Einstein formula gives
\begin{equation}
\frac{D_{e}^{ij}}{D} = \delta_{ij} +  \int_{\Omega_{||}} d{\bf x} \left[ f_j\left({\bf x},h_2({\bf x})\right) \nabla_{||i}h_2({\bf x}) -
f_j\left({\bf x},h_1({\bf x})\right) \nabla_{||i}h_1({\bf x}) \right].\label{kubo1}
\end{equation}

Differentiating the steady state equation (\ref{ss}) with respect to $u_{||i}$ and then setting $u_{||i}=0$ yields the equation obeyed by the vector field $f_i$:
\begin{equation}
\nabla^2 f_i =0. \label{lap}
\end{equation}
Furthermore, $ f_i({\bf x},y)$ must be periodic in the in plane coordinates ${\bf x}$ and the normalization of the steady state distribution over $\Omega$ implies that
\begin{equation}
\int_\Omega dyd{\bf x}\ f_i({\bf x},y) = 0,\label{int}
\end{equation}
while the no flux boundary conditions at the channel surface imposes 
\begin{equation}
{\bf n_\alpha}\cdot\left[-\nabla f_i\left({\bf x}, y\right)|_{y=h_\alpha({\bf x})} +\frac{{\bf e}_i}{\Omega}\right]=0,\label{bcs}
\end{equation}
where ${\bf e}_i$ denotes the unit vector in the direction $i$ in the plane parallel to the channel. The above condition can be rewritten
as 
\begin{equation}
-\nabla_{||}h_\alpha ({\bf x})\cdot\nabla_{||}f_i\left({\bf x}, y\right)|_{y=h_\alpha({\bf x})} + \frac{\partial f_i}{\partial y}({\bf x},h_\alpha({\bf x}))= -\frac{\nabla_{||i}h_\alpha({\bf x})}{\Omega}.\label{bcs2}
\end{equation}
To summarize, an exact formula for the effective diffusivity is given by Eq.~(\ref{kubo1}), where the auxiliary function $f_i$ is computed by solving Laplace's equation (\ref{lap}) with boundary conditions (\ref{bcs2}). The above results are in full accordance with the general theory of Refs.~\cite{gue2015a,gue2015b} and with the Brenner macro transport theory \cite{bre1993}.

\section{Formulation of the problem with analytic functions for two dimensional symmetric channels}
\label{Sec2}

 Most studies in the literature have considered the case where the channel is two dimensional.
 Here we will denote the 
coordinate along the channel as $x$ and, as earlier, use  $y$ to denote the height coordinate. We will consider channels which are symmetric about the axis
$y=0$, that is to say we write $h(x)=h_2(x)=-h_1(x)$. We define $L$ as the period of the channel, so that $h(x+L)=h(x)$. 

Here there is only one variable $f_1=f$ corresponding to the $x$ direction it obeys
Laplace's equation 
\begin{equation}
\nabla^2 f(x,y)=0.
\end{equation}
The solution of Laplace's equation can be written using the theory of analytic functions as the real  or  imaginary part (or a linear combination of both) of
\begin{equation}
f(x,y) = W_1(z) + W_2(\overline z),
\end{equation}
where $z= x+iy$ and $\overline z= x-iy$ denotes its complex conjugate. For symmetric channels, the  symmetry about $y=0$ implies that $f(x,y)=f(x,-y)$ and thus one can easily see that
\begin{equation}
W_1(z) = W_2(z) + c,
\end{equation}
where $c$ is a constant. However, $c$ can simply be incorporated into the definition of $W_1$ so we can use the form
\begin{equation}
f(x,y) = W_1(z) + W_1(\overline z).
\end{equation}
Since $f(x,y)$ is real, we can assume without loss of generality that the coefficients of the Laurent expansion of $W_1$ are real, so that $\overline{ W_{1}(z)}= W_{1}(\overline z)$. Making a convenient rescaling in terms of the unit channel area $\Omega$ we thus see that, for a symmetric channel, one can look for solutions of the form 
\begin{equation}
f(x,y) = \frac{1}{2\Omega}[w(x+iy) + w(x-iy)],\label{gen}
\end{equation}
where $w$ is analytic such that $\overline{ w(z)}= w(\overline z)$ and is periodic in $x$ (with period $L$). If we write $w$ in terms of its real and imaginary parts, $w(x+iy) = u(x,y) +iv(x,y)$, the symmetry $y\to-y$ implies that
\begin{equation}
u(x,-y) = u(x,y)
\end{equation}
This means, looking at an expansion in terms of $y$ at fixed $x$, that $\partial u/\partial y$ is an odd function in $y$. However, the Cauchy-Riemann equation
\begin{equation}
\frac{\partial u}{\partial y}(x,y) = -\frac{\partial v}{\partial x}(x,y),\label{CR}
\end{equation}
shows that $\partial v/\partial x$ is odd in $y$. However, at $y=0$ this equation implies, as $\frac{\partial u}{\partial y}|_{y=0} =0$, that $v(x,0)$ is a constant ($v_0$). Consequently, we have, integrating Eq.~(\ref{CR}) over $x$, that
\begin{equation}
v(x,y) = v_0 + V(x,y)\label{vo}
\end{equation}
where $V(x,-y)=-V(x,y)$. However there is nothing in the problem that allows the  determination of  $v_0$, furthermore one can show that it does not affect the value of the diffusion constant, and we thus chose $v_0=0$ and so $v(x,-y)=-v(x,y)$. 

Inserting the ansatz in Eq.~(\ref{gen})  into the boundary condition (\ref{bcs2}) provides an additional condition for the value of $w$ at the channel boundary:
\begin{equation}
ih'(x)[w'\left(x+ih(x))+w'(x-ih(x)\right)] +[w'\left(x+ih(x)\right)-w'\left(x-ih(x)\right)] = i2{h'(x)},
\label{bcc}
\end{equation}
and this can be integrated with respect to $x$ to yield
\begin{equation}
w\left(x+ih(x)\right) - w\left(x-ih(x)\right) = 2ih(x)-i2C,\label{be}
\end{equation}
where $C$ is a constant to be determined. Finally, the effective diffusivity is given by
\begin{equation}
\frac{D_e}{D} = 1 +  2\int_0^L d{x} f({x},h({ x})) h'({x}), \label{kubo}
\end{equation}
and so within this  formalism, the diffusion constant is simply given by
\begin{eqnarray}
\frac{D_e}{D} &=& 1 +  \frac{1}{\Omega}\int_0^L d{x} [w\left(x+ih(x)\right)+w\left(x-ih(x)\right)] h'({x}).\label{kubo2}
\end{eqnarray}
However, we show in appendix \ref{AppendixConstantC} that the above equation can be simplified, leading to
\begin{eqnarray}
\frac{D_e}{D} &=& \frac{C}{\langle h \rangle_s}, \label{kubo2WithConstantC} ,
\end{eqnarray}
where 
\begin{equation}
{\langle h \rangle_s}= \frac{1}{L}\int_0^L dx \ h(x),
\end{equation}
is the channel height spatially averaged over one period.
In the following section we present an alternative proof based on a perturbation expansion.

We have thus reformulated the diffusivity problem for two-dimensional symmetric channels into a problem of finding an analytic function $w(z)$ in the complex plane $z=x+iy$, with the condition (\ref{be}) at the channel boundary. Remarkably, once the constant $C$ is determined in (\ref{be}), there is no more integration to be done to determine $D_e$ which is directly proportional to this constant [ see Eq.~(\ref{kubo2WithConstantC})]. The use of analytic functions to analyze FJ type problems in two dimensional channels has of course been exploited in a number of studies \cite{kal2010,kal2014,kal2015}; however, the present study shows that this method can be exploited much further.  

\section{Perturbation expansion for the effective diffusivity}
\label{Sec3}

\subsection{Recovering the basic Fick-Jacobs' approximation}
We now show how our complex formalism can be used to recover the basic FJ approximation for the effective diffusivity.
To proceed we assume that the length of the channel is much larger than the height and thus expand Eq.~({\ref{be}) for small
$h$ which gives
\begin{equation}
w'(x)h(x) = h(x) - C,
\end{equation}
Integrating and using the periodicity of $w$ we find
\begin{equation}
w(x) = \left[x- \langle h^{-1}\rangle^{-1}_s\int_0^x \frac{dx'}{h(x')}\right] + C'
\end{equation}
where 
\begin{equation}
\langle h^{-1}\rangle_s = \frac{1}{L}\int _0^L \frac{dx'}{h(x')},
\end{equation}
is the spatial average of $h^{-1}$ over the unit cell. The term $C'$ is determined from the normalization condition
but we do not need to evaluate it gives zero contribution to the effective diffusivity as can be easily seen in Eq.~(\ref{kubo2}). 
Keeping the leading order in $h$ in Eq.~(\ref{kubo2}) now gives
\begin{equation}
\frac{D_e}{D}  \simeq 1 - \frac{2}{\Omega} \int_0^L d{x} \ w'({x})h(x).
\end{equation}
Now, noting that $\Omega = 2\int_0^L dx\ h(x)$ we find
\begin{equation}
\frac{D_e}{D} \approx \frac{D_{\rm FJ}}{D} = \frac{1}{\langle h^{-1}\rangle_s\langle h\rangle_s},\label{fj}
\end{equation}
this is exactly the same result as given by the basic FJ formula (that is to say with no modifications to the local diffusion constant). 

\subsection{Perturbation theory in channel width}\label{pert}

We now consider a channel height of the form 
\begin{align}
h(x) = \varepsilon\zeta(x), 
\end{align}
which we will analyze here in the limit of small $\varepsilon$.
We now show that our complex formalism can be systematically improved to obtain an expansion of the effective diffusivity in powers of the small parameter $\varepsilon$ in a very compact way.
 
The whole perturbation analysis for the function $w(x)$ is based on Eq.~(\ref{be}) which can be written as
\begin{equation}
\sum_{n=0}^\infty(-1)^n\varepsilon^{2n+1}\frac{ w^{(2n+1)}(x) \zeta^{2n+1} (x)}{(2n+1)!} = {\varepsilon \zeta(x)-C},\label{master}
\end{equation} 
where we use the notation $w^{(n)} = d^n w /dx^n$. The above equation is compatible with an expansion of $w$ in even powers of $\varepsilon$ of the form
\begin{equation}
w(x) = \sum_{m=0}^\infty \varepsilon^{2m}w_{2m}(x),
\end{equation}
and the constant $C$ of the form
\begin{equation}
C= \varepsilon \sum_{k=0}^\infty \varepsilon^{2k}C_{2k} .
\end{equation}
The boundary condition Eq.~(\ref{be}), expressed as a perturbation series in $\varepsilon$, now becomes for $k>0$,
\begin{equation}
\sum_{n,m=0}^\infty(-1)^n\varepsilon^{2n+2m}\frac{ w_{2m}^{(2n+1)}(x) \zeta^{2n+1} (x)}{(2n+1)!} = { \zeta(x)-\sum_{k=0}^\infty \varepsilon^{2k}C_{2k} }.\label{perta}
\end{equation}

We can compute directly using  Eq.~(\ref{kubo}), the expansion in $\varepsilon$ of the effective diffusivity, finding
\begin{equation}
\frac{D_e}{D} =1 - \frac{1}{\langle \zeta\rangle_s}\int_0^L d{x} \sum_{n=0}^\infty (-1)^n \varepsilon^{2n}\frac{w^{(2n+1)}(x)\zeta^{2n+1}(x)}{(2n+1)!},
\end{equation}
after an integration by parts. It follows from the relation Eq. (\ref{master}) that the constants $C_i$ directly provide the terms of the perturbative expansion of the effective diffusion constant, which reads
\begin{equation}
\frac{D_e}{D} =\frac{C}{\varepsilon\langle \zeta\rangle_s} = \frac{1}{\langle \zeta\rangle_s} \sum_{k=0}^\infty C_{2k} \varepsilon^{2k} \label{De},
\end{equation}
thus providing an alternative derivation of Eq.~(\ref{kubo2WithConstantC}) to that derived in appendix \ref{AppendixConstantC}.

Let us write explicitly the first orders in $\varepsilon^k$ of Eq.~(\ref{perta}). The leading order term in $\varepsilon$ gives the basic FJ approximation found above
\begin{equation}
\zeta(x)w_0^{(1)} (x) = \zeta(x)-C_0,\label{ord0}
\end{equation}
while the higher order corrections give for $k>0$,
\begin{equation}
\sum_{n=0}^k (-1)^n\frac{ w_{2(k-n)}^{(2n+1)}(x) \zeta^{2n+1} (x)}{(2n+1)!} = -{C_{2k}}. \label{09431}
\end{equation}
The next three orders in perturbation theory provide the equations satisfied by $w_2,w_4,w_6$ and read:
\begin{equation}
-\frac{1}{3!}w_0^{(3)}(x) \zeta^{3}(x) + w_2^{(1)}(x) \zeta(x)= -{C_2},\label{ord2}
\end{equation}
for $k=1$,
\begin{equation}
\frac{1}{5!}w_0^{(5)}(x) \zeta^{5}(x) - \frac{1}{3!}w_2^{(3)}(x) \zeta^3(x)+ w_4^{(1)}(x) \zeta(x) = -{C_4}, \label{ord4} 
\end{equation}
for $k=2$ and
\begin{equation}
- \frac{1}{7!} w_0^{(7)}(x)\zeta(x)^7 + \frac{1}{5!} w_2^{(5)}(x) \zeta(x)^5 -\frac{1}{3!} w_4^{(3)}(x) \zeta(x)^3 + w_6^{(1)}(x) \zeta(x)  =-C_6, \label{ord6} 
\end{equation}
for $k=3$.

We now show how to compute the constants $C_{2k}$. For $k=0$, the value of $C_0$ is found by 
imposing periodic boundary conditions in Eq.~(\ref{ord0}), which gives 
\begin{align}
C_0 = \langle \zeta^{-1}\rangle_s^{-1}.
\end{align}
Inserting into Eq.~(\ref{De}) one sees that one of course recovers the basic form of the FJ approximation, the solution for $w_0'$ is in this notation
\begin{equation}
w_0^{(1)}(x) = 1-  \frac{1}{\langle \zeta^{-1}\rangle_s\zeta(x)}
\end{equation}
Substituting this solution into Eq.~(\ref{ord2}) now gives
\begin{equation}
w_2^{(1)}(x) = -\frac{C_2}{\zeta(x)} +\frac{\langle \zeta^{-1}\rangle^{-1}_s}{6}\left[\zeta^{(2)}(x)- 2 \frac{\zeta^{(1)}(x)^2}{\zeta(x)}\right]
\end{equation}
and the periodicity condition $w_2(0)=w_2(1)$ now gives
\begin{equation}
C_2 = -\frac{1}{3}\frac{\langle \zeta'^2/\zeta\rangle_s}{\langle \zeta^{-1}\rangle_s^2} .
\end{equation}
Substituting the solution for $w^{(1)}_2$ into Eq.~(\ref{ord4}) yields
\begin{equation}
w_4^{(1)}(x) = -\frac{C_4}{\zeta(x)} +\frac{1}{6}\zeta^2(x) \left[ -\frac{C_2}{\zeta(x)} +\frac{C_0}{6}\left(\zeta^{(2)}(x)- 2 \frac{\zeta^{(1)}(x)^2}{\zeta(x)}\right)\right]^{(2)}+\frac{C_0}{120}\zeta^{4}(x)\left[\frac{1}{\zeta(x)}\right]^{(4)}.
\end{equation}
The periodicity of $w_4$ then gives, after some integration by parts, 
\begin{equation}
\frac{C_4}{C_0} = \frac{C_2^2}{C_0^2} + \frac{C_0}{45} \left[ 4 \langle \zeta'^4/\zeta \rangle_s +  \langle \zeta''^2\zeta\rangle_s \right] ,
\end{equation}
which can be rearranged as
\begin{equation}
C_4 = \frac{\langle \zeta'^2/\zeta\rangle_s^2}{9 \langle \zeta^{-1} \rangle_s^3} + \frac{1}{45} \frac{4 \langle \zeta'^4/\zeta \rangle_s +  \langle \zeta''^2\zeta\rangle_s}{\langle \zeta^{-1} \rangle_s^2}.
\end{equation}

Using the above expressions for $C_0$, $C_2$ et $C_4$ we find to $O(\varepsilon^4$):
\begin{equation}
\frac{D_e}{D} = \frac{1}{\langle \zeta \rangle_s \langle \zeta^{-1} \rangle_s} \left[ 1 - \varepsilon^2 \frac{\langle \zeta'^2/\zeta\rangle_s}{3 \langle \zeta^{-1} \rangle_s} +\varepsilon^4 \left( \frac{\langle \zeta'^2/\zeta\rangle_s^2}{9 \langle \zeta^{-1} \rangle_s^2} + \frac{4 \langle \zeta'^4/\zeta \rangle_s }{45 \langle \zeta^{-1} \rangle_s} + \frac{ \langle \zeta''^2\zeta\rangle_s}{45 \langle \zeta^{-1} \rangle_s} \right)\right].\label{o4}
\end{equation}
This agrees with the results of Refs.~\cite{kal2006,dorfman2014assessing}. In Ref.~\cite{kal2006} the result was obtained by a formalism which derives an effective one dimensional diffusion constant. However, as pointed out before, the reduction to an effective one dimensional problem is not obvious as the resulting one dimensional process is non-Markovian. In Ref.~\cite{dorfman2014assessing} the effective macroscopic transport theory of Brenner \cite{bre1993} was used. 
The expression for the diffusion constant in this theory can actually be shown to agree with the Kubo formula derived here. An interesting point made in Ref.~\cite{dorfman2014assessing} is that the result of Ref.~\cite{kal2006} is recovered at orders up to and including $\varepsilon^4$. 

Here we exploit the compact nature of the analysis here to derive the
term of $\mathcal{O}(\varepsilon^6)$ and show that it still agrees with Ref.~\cite{kal2006}.
Substituting the solution for $w^{(1)}_0$, $w^{(1)}_2$ and $w^{(1)}_4$ into Eq.~(\ref{ord6}) and imposing the periodicity of $w_6$ gives the value of $C_6$ as
\begin{equation}
\begin{split}
\frac{C_6}{C_0}=&-\frac{C_4}{3} \langle \zeta'^2/\zeta \rangle_s +\frac{C_2}{45} \langle 4\zeta'^4/\zeta +\zeta''^2\zeta \rangle_s \\
&- \frac{C_0}{945} \langle 44 \zeta'^6/\zeta +5 \zeta^2\zeta''^3 -60 \zeta'^4 \zeta'' +9 \zeta\zeta'^2\zeta''^2 -4 \zeta\zeta'^3\zeta'''+2 \zeta^2\zeta'\zeta''\zeta'''-2\zeta^3\zeta''\zeta^{(4)}+4\zeta^2\zeta'^2\zeta^{(4)} \rangle_s.
\end{split}
\end{equation}
Using integration by parts to minimize the highest derivative of $\zeta$ appearing in the final expressions we obtain
\begin{equation}
\frac{C_6}{C_0}= \frac{C_2^3}{C_0^3} + \frac{2C_4C_2}{C_0^2} - \frac{C_0}{945} \left[ 44 \langle \zeta'^6/\zeta \rangle_s+5 \langle\zeta^2\zeta''^3 \rangle_s+45 \langle \zeta'^3\zeta''^2\rangle_s +2 \langle\zeta^3\zeta'''^2\rangle_s \right],
\end{equation}
which can be rearranged as
\begin{equation}
C_6= -\frac{1}{27} \frac{\langle \zeta'^2/\zeta\rangle_s^3}{ \langle \zeta^{-1} \rangle_s^4} - \frac{2}{135} \frac{\langle \zeta'^2/\zeta \rangle_s \langle 4\zeta'^4/\zeta +\zeta''^2\zeta \rangle_s}{\langle \zeta^{-1} \rangle_s^3} - \frac{1}{945} \frac{ 44 \langle \zeta'^6/\zeta \rangle_s+5 \langle\zeta^2\zeta''^3 \rangle_s+45 \langle \zeta'^3\zeta''^2\rangle_s +2 \langle\zeta^3\zeta'''^2\rangle_s }{\langle \zeta^{-1} \rangle_s^2}
\end{equation}
The final expression for the diffusion constant to $O(\varepsilon^6)$ can also be conveniently written as
\begin{eqnarray}
\frac{D_e}{D} &=& \langle \zeta \rangle_s^{-1} \left\{\langle \zeta^{-1} \rangle_s + \frac{\varepsilon^2}{3}\langle \zeta'^2/\zeta\rangle_s-\frac{\varepsilon^4}{45} \left[4 \langle \zeta'^4/\zeta \rangle_s +\langle \zeta''^2\zeta \rangle_s\right] \right. \nonumber \\
 &+& 
\left.\frac{\varepsilon^6}{945} \left[44 \langle \zeta'^6/\zeta \rangle_s+5 \langle\zeta^2\zeta''^3 \rangle_s+45 \langle \zeta'^3\zeta''^2\rangle_s +2 \langle\zeta^3\zeta'''^2\rangle_s \right] \right\}^{-1},
\end{eqnarray}
which is also true at order $\mathcal{O}(\varepsilon^6)$ and after some work one can show that this agrees with the result of \cite{kal2006} at $\mathcal{O}(\varepsilon^6)$. 

\section{The pseudo-linear channel approximation}
\label{Sec4}

In Ref.~\cite{kal2006} the perturbation theory was partially resumed by assuming that derivatives $\zeta''$ and of higher order vanish. Here we show how this re summation can be derived within the formalism developed here in a very compact manner. We start from the Eq.~(\ref{be}) and make the ansatz
\begin{equation}
w(x+iy) = x +iy -\frac{1}{\langle\phi\rangle_s}\int_0^{x+iy} d\xi \phi(\xi),
\end{equation}
where 
\begin{equation}
\langle\phi\rangle_s = \int_0^1 \phi(x) dx
\end{equation}
is the integral of $\phi$ along the line $y=0$. In addition, we assume that $\phi$ is analytic and so the integral in the second term on the right hand side is independent of the integration path between $0$ and $x+iy$. We also assume that $\phi$ is of period $1$ in the variable $x$, {\em i.e.} $\phi(x+1+iy) = \phi(x+iy)$, so we are working in units where the period $L=1$ to simplify notation.

We now note that the function $w$ so constructed is analytic and is periodic in $x$ with period $1$. To see this latter point we write
\begin{equation}
w(x+1+iy) = x +1 + iy -\frac{1}{\langle\phi\rangle_s}\int_0^{x+1+iy} d\xi \phi(\xi).\label{04532}
\end{equation}
The contour integral on the right hand side can be decomposed into an integral along 
the real axis between $0$ and $x+1$ followed by an integral parallel to the imaginary axis
between $x+1$ and $x+1+iy$ so we can write 
\begin{equation}
w(x+1+iy) = x +1 + iy -\frac{1}{\langle\phi\rangle_s}\int_0^{x+1} dx'\phi(x')
- \frac{i}{\langle\phi\rangle_s}\int_0^{y} dy' \phi(x+1 + iy')
\end{equation}
However, by the periodicity of $\phi$ we can write
\begin{equation}
\frac{1}{\langle\phi\rangle_s}\int_0^{x+1} dx'\phi(x') = 1 + \frac{1}{\langle\phi\rangle_s}\int_0^{x}
dx'\phi(x')
\end{equation}
and 
\begin{equation}
\int_0^{y} dy' \phi(x+1 + iy') = \int_0^{y} dy' \phi(x + iy'),
\end{equation}
which once inserted into (\ref{04532}) proves the desired periodicity, {\em i.e.} $w(x+1+iy)=w(x+iy)$. We now substitute this ansatz into the boundary condition (\ref{be}) which becomes
\begin{equation}
\frac{1}{\langle \phi\rangle_s}\left[ \int_0^{x+ih(x)} d\xi \phi(\xi) - \int_0^{x-ih(x)} d\xi \phi(\xi)\right] = 2iC\label{eqcomp}
\end{equation}
We now note that the overall contour integral on the left hand side goes over the line between
$x-ih(x)$ and $0$ then the line from $0$ to $x+ih(x)$. Using the analytic property of $\phi$ we deform the contour to the straight line between $x-ih(x)$ and $x+ih(x)$ which is parallel to the imaginary axis so we can write
\begin{equation}
\frac{1}{\langle \phi\rangle_s}\int_{-1}^1 ds \ h(x) \ \phi\left(x + ish(x)\right) = 2C.
\end{equation}
It is easy to verify that if we write $h(x) = \varepsilon\zeta(x)$ the above generates the same 
perturbation theory as was found in the previous section. However, the starting point of the perturbation theory can be changed by writing
\begin{equation}
\phi(\xi) = \frac{\psi(\xi)}{h(\xi)},
\end{equation}
which leads to 
\begin{equation}
\frac{1}{\langle \frac{\psi}{h}\rangle_s}\int_{-1}^1 ds\ \zeta(x) \ \frac{\psi\left(x + is\varepsilon\zeta(x)\right)}{
\zeta\left(x + is\varepsilon\zeta(x)\right)} = 2C.\label{pla1}
\end{equation}
At lowest order in $\varepsilon$ it is easy to verify using Eq.~(\ref{De}) that this recovers the basic FJ approximation as the solution is $\psi(x) = \psi_0$ where $\psi_0$ is constant.

To recover the pseudo-linear approximation of \cite{kal2006}, we proceed as follows. First, we ignore the terms $\zeta''(x)$ and higher orders in Taylor expanding the denominator of the integrand on the right hand side of Eq.~(\ref{pla1}), so that we approximate
\begin{align}
\zeta(x+i s \varepsilon \zeta(x))\simeq \zeta(x)+\zeta'(x) \ i s \varepsilon \zeta(x)
\end{align}
Next, we also write 
\begin{align}
\psi\left(x + is\varepsilon\zeta(x)\right)\approx\psi(x), \label{ApproxPsi}
\end{align}
later we will verify that this is consistent with neglecting terms $h''$ and higher derivatives. With these approximations, Eq.~(\ref{pla1}) becomes
\begin{equation}
\frac{\psi(x)}{\langle \frac{\psi}{h}\rangle_s} \int_{-1}^1 \frac{ds}{1+i s \varepsilon \zeta'(x)} =2C .
\end{equation}
Performing the integral in the above expression leads to
 \begin{equation}
\psi(x) = C\langle \frac{\psi}{h}\rangle_s \frac{\varepsilon |\zeta'(x)|}{\tan^{-1}(\varepsilon |\zeta'(x)|)}.
\label{pla2}
\end{equation}
where $\mathrm{tan}^{-1}$ is the arctangent function. Here we see that $\psi'(x)\sim h''(x)$ and thus the approximation made in (\ref{ApproxPsi}) is consistent. Finally,
multiplying both sides of Eq.~(\ref{pla2}) by $1/h=1/(\varepsilon\zeta)$ and integrating over $[0,1]$ we find
\begin{equation}
C = \left\langle\frac{ |h'(x)|}{h(x)\tan^{-1}( |h'(x)|)}\right\rangle_s^{-1},
\end{equation}
and thus Eq.~(\ref{kubo2WithConstantC}) gives
\begin{equation}
\frac{D_e}{D} = \frac{1}{\langle h \rangle_s \langle\frac{ h'(x)}{h(x)\tan^{-1}(h'(x))}\rangle_s}.\label{pwl}
\end{equation}

Within their systematic reduction to a one dimensional diffusion equation the authors of \cite{kal2006} show that the effective one dimensional diffusion constant that should be used in
Eq.~(\ref{lif}) is 
\begin{equation}
D(x) = D \frac{\tan^{-1}\left(h'(x)\right)}{h'(x)},
\end{equation}
and hence by using the Lifson-Jackson formula (\ref{lif}) we recover exactly the same formula (\ref{pwl}) under the same approximation.

\section{Generating Exact results}
\label{Sec5}
Here we re derive in a very direct manner a classical exact result due to Zwanzig \cite{zwa1982}
for channels which can be appropriately parameterized in another coordinate system.
Consider the transformation
\begin{equation}
z = G(w)
\end{equation}
where $z = x+iy$ and $w=u+iv$ where $G$ is an analytic function.  Importantly, the upper and lower channel boundaries are defined to be given by $v=\pm V$ where $V$ is constant. Of course 
the function $G$ must be chosen to be periodic in $u$ and should be single valued. Equation 
(\ref{eqcomp}) is trivially written as
\begin{equation}
\frac{1}{\langle \phi\rangle_s}\int^{x+ih(x)}_{x-ih(x)} d\xi\ \phi(\xi) = 2iC\label{eqcomp2},
\end{equation}
where the integral is a contour integral over any finite contour joining the lower to upper limits indicated on the integral. Now in terms of the variables $w$ this becomes
\begin{equation}
\frac{1}{\langle \phi\rangle_s}\int^{u+iV}_{u-iV} dw \frac{dG}{dw} \phi(w) = 2iC.
\end{equation}
The solution to this equation (up to a multiplicative constant) is now clearly given by
\begin{equation}
\phi(w) = \left[\frac{dG}{dw}\right]^{-1},
\end{equation}
which gives
\begin{equation}
C = \frac{V}{\left\langle\left[\frac{dG}{dw}\right]^{-1}\right\rangle_s},
\end{equation}
and Eq.~(\ref{kubo2WithConstantC}) now becomes
\begin{equation}
\frac{D_e}{D}=\frac{V}{\left\langle\left[\frac{dG}{dw}\right]^{-1}\right\rangle_s\langle h\rangle_s}.
\end{equation}
This can be rewritten by noticing that $\langle h\rangle_s=\Omega/2$ is half the area of the unit cell and so we can write in the coordinate system given by $w$ 
\begin{equation}
\langle h\rangle_s=\frac{1}{2}\int_0^1 du \int_{-V}^{V} dv \left|\frac{dG}{dw}\right|^2.
\end{equation}
For a symmetric channel we can make the line $y=0$ coincide with the line $v=0$, so $G$ is real for $v=0$, 
\begin{equation}
\left\langle\left[\frac{dG}{dw}\right]^{-1}\right\rangle_s= \int_0^1 dx \left[\left.\frac{dG}{dw}\right|_{v=0}\right]^{-1} = \int_0^1 du G'(u)[G'(u)]^{-1} = 1,
\end{equation}
where we have used the notation $\frac{dG}{dw}|_{v=0} = G'(u)$. Putting this altogether then gives 
\begin{equation}
\frac{D_e}{D}=2\frac{V}{\int_0^1 du \int_{-V}^{V} dv |\frac{dG}{dw}|^2}. \label{exactzwanzig}
\end{equation}
Up to differences in notation the above is identical to the result of Zwanzig \cite{zwa1982}.

\section{The large $\varepsilon$ limit}\label{largee}
\label{Sec6}
Here we can consider the case of very wide channels where we again write $h(x)=\varepsilon\zeta(x)$ but we assume that $\varepsilon\to\infty$. Before analyzing this problem via the formalism developed above 
we propose a simple physical argument which gives a bound on the diffusion constant. 
Consider a symmetric channel whose height function about the center of the channel achieves its minimum value at some point $x_{\rm min}$ where it takes the value $h_{\rm min}$. Without loss of generality we choose $x_{\rm min}=-1/2\equiv 1/2$. Now while the particle is in the region
$\Omega' = [-1/2,1/2]\times[-h_{\rm min},h_{\rm min}]$ the particle diffuses without hindrance from the sides of the channel, so we can write
\begin{equation}
\frac{d}{dt}\langle X^2 \rangle = 2D,
\end{equation}
however, outside of this region the diffusion is slowed down. If we denote by $t_{\Omega'}$ the total time spent in  $\Omega'$ we can thus write
\begin{equation}
\langle X^2 \rangle \geq 2Dt_{\Omega'},
\end{equation}
because time spent in the region where the walls can interfere with the diffusion will nonetheless 
contribute to diffusion - for example consider small excursions into this region which come back into $\Omega'$ having moved along the channel.

As we know the equilibrium distribution in the channel modulo the period is uniform, ergodicity then implies that
\begin{equation}
\frac{t_{\Omega'}}{t}=\frac{|\Omega'|}{|\Omega|} = \frac{h_{\rm min}}{\langle h\rangle_s}
\end{equation}
which gives the bound
\begin{equation}
\langle X^2 \rangle \geq 2D\frac{h_{\rm min}}{\langle h\rangle_s}t
\end{equation}
and thus we have the bound on the effective diffusion constant along the channel 
\begin{equation}
\frac{D_e}{D} \geq \frac{h_{\rm min}}{\langle h\rangle_s} = \frac{\zeta_{\rm min}}{\langle \zeta\rangle_s}.\label{in1}
\end{equation}
It is also clear that $D_e \le D$ and thus we have the two sided bound
\begin{equation}
1\geq \frac{D_e}{D} \geq \frac{\zeta_{\rm min}}{\langle \zeta\rangle_s},\label{bound}
\end{equation}
and this simple result shows one that the system has a finite diffusion coefficient even in the limit $\varepsilon\to \infty$. 
In \cite{dea1993} diffusion on comb-like locally one dimensional structures was studied. The comb consists of a one dimensional backbone along which the diffusion is measured with teeth perpendicular to the backbone, excursions into the teeth slow down diffusion along the background. In this case the lower bound in Eq.~(\ref{bound}) becomes exact as no diffusion can occur in the teeth due to their one dimensional character and given they are perpendicular to the backbone. Other problems of biased diffusion in tubes with dead-ends are studied in \cite{goo1960,dag2007,ber2011b}.

\begin{figure}
\begin{center}
  \includegraphics[scale=0.45]{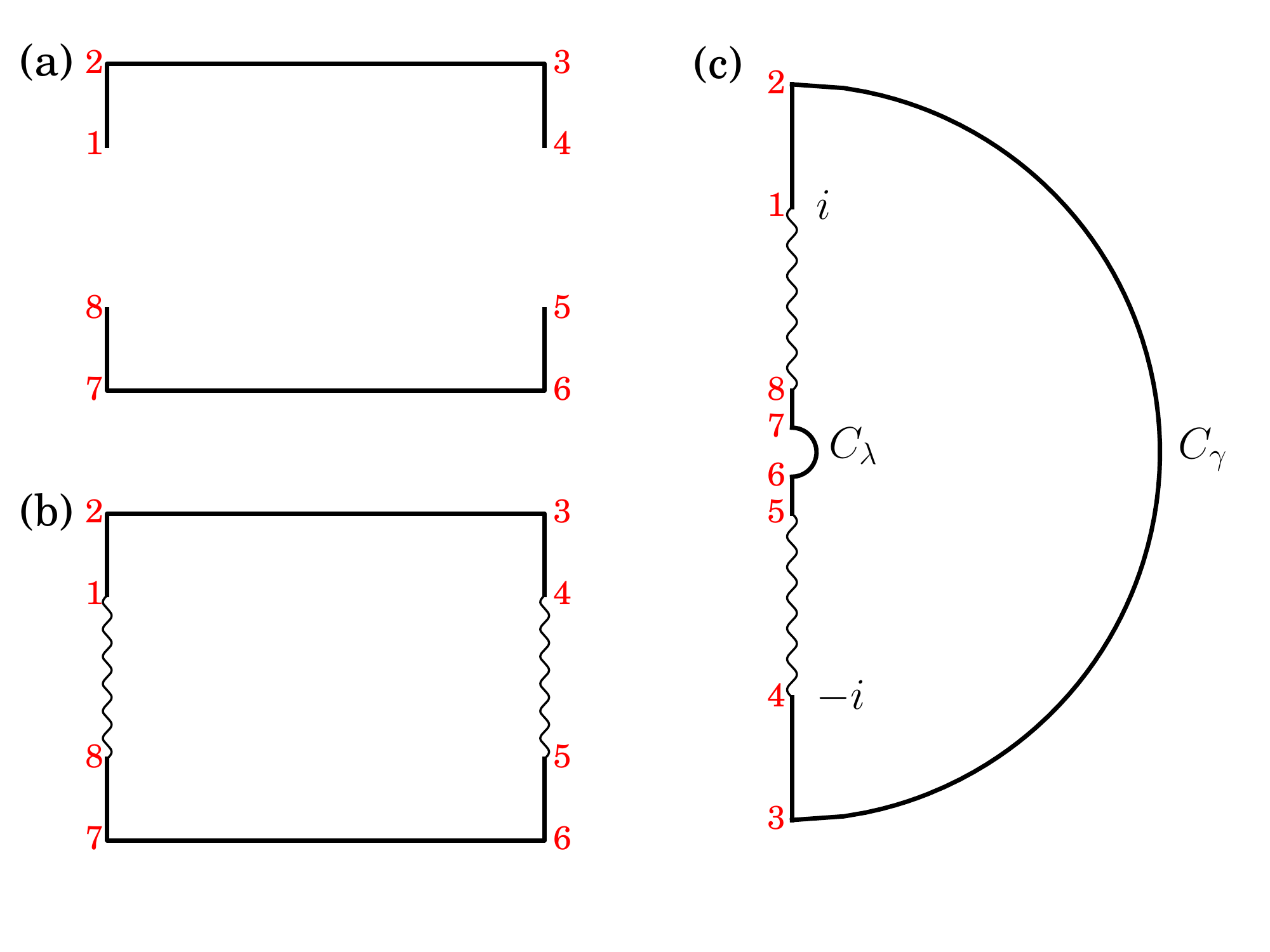}
  \caption{\label{conform}(color online) (a) The domain of the analytic function $w(z)$. (b) The domain of the analytic function $\tilde w(z) = w(z)-z$. The wiggly lines represents a jump of +1 in the solution. (c) The domain of the analytic function $\tilde w(\xi)$, where $\xi=\exp \left[-i\pi (z-ih_{\rm min})\right]$. A jump of +1 in the solution occurs between the two curved lines.}
\end{center}
\end{figure}
 
To go beyond this analysis we consider a channel where the points $x=\pm 1/2$ correspond to the minimum of the height of the channel. The height at $x=\pm1/2$ is, as above, denoted by $h_{\rm min}$. 
To start with we define a function $\tilde w(z)$ via
\begin{equation}
w(z) = z+\tilde w(z),
\end{equation}
where $w(z)$ is the solution sought in Eq.~(\ref{gen}). The boundary Eq.~(\ref{be}) contain now becomes 
\begin{equation}
\tilde w\left(x+ih(x)\right) - \tilde w\left(x-ih(x)\right) = -i2C\label{be2},
\end{equation}
however, the function $\tilde w$ is no longer periodic and we find
\begin{equation}
\tilde w(x+1 +iy) -\tilde w(x+iy) = -1.\label{jump}
\end{equation}
Now we make the transformation $\xi=\exp \left[-i \pi (z-ih_{\rm min}) \right]$. This conformal mapping is made explicit in the simple case where $\Omega$ is the  septate channel shown in Fig.~\ref{conform} and  where we denote by $h_\mathrm{max}$ the maximal channel height. The domain shown in Fig.~\ref{conform}a also represents the region in which the original function $w(z)$ is analytic. In Fig.~\ref{conform}b we show the same region but include the branch cuts over which the function $\tilde w(z)$ jumps. In this case the domain $\Omega$ is mapped into that shown on the Fig.~\ref{conform}c, where the radius of the outer half-circle $C_\gamma$ is given by $\gamma=\exp \left[{\pi} (h_{\rm max}-h_{\rm min})\right]$ and that of the inner circle $C_\lambda$ is $\lambda=\exp \left[-\pi (h_{\rm max}+h_{\rm min})\right]$. The crucial point is that in the limit $\varepsilon \to \infty$ the radius $\gamma$ becomes infinite while the radius $\lambda$ goes to zero (we recall that $h$ is  proportional to $\varepsilon$). For a general channel the points on the upper surface are mapped to a point at infinity while those on the lower surface are mapped to $0$. We therefore argue that the obtained results will be valid for large $\varepsilon$ for arbitrary channel shapes, since the details of the shape profile are sent to infinity after the mapping to the $\xi$ space. 

The resulting Laplace problem on the new domain corresponds to a classic two dimensional electrostatic problem, one has  two charged plates, where the potential is $\phi=1/2$ on an upper plate, between $(0,0)$ and $(0,1)$, and $\phi=-1/2$ on the lower plate, between $(0,0)$ and $(0,-1)$. The potential vanishes at infinity. The solution is given by $\tilde w(\xi)=\frac{i}{\pi}\rm{arcsinh}(1/\xi)$. 
The solution $w(z)$ found here can include an arbitrary, undetermined, imaginary constant, corresponding to the unimportant constant $v_0$ in Eq.~(\ref{vo}). Taking the logarithmic representation of $\rm{arcsinh}$,
we find
\begin{equation}
w(x+iy)=\frac{i}{\pi} \ln \left[1+\sqrt{1+\exp \left(-{2\pi}ix + {2\pi}(y-h_{\rm min})\right)}\right] +i\tilde C.\label{explicit}
\end{equation}
where $\tilde C$ is a constant. We note that, for $y\ll h_{\rm min}$ and $h_{\rm min} \gg 1$, we have $w(x)= i \ln(2) / \pi + i\tilde C$, and we recall  that the choice $v_0=0$ means that the imaginary part vanishes on the central line $y=0$, this fixes the value of $\tilde C=- \ln(2) /\pi $. The above equation is valid for $y>0$. For $y<0$ the solution can be  determined  by symmetry to be $w(\overline z)=\overline w(z)$. 

Next we note that $w(x+iy)-w(x-iy)$ in Eq.(\ref{be}) is simply $2i\text{Im}[w(x+iy)]$, so that we can rewrite this equation and identify the constant $C$, considering the limit $h(x) \gg 1$,
\begin{align}
C= h_{\rm min} + \frac{\ln 2}{\pi}
\end{align}
and finally we use Eq.~(\ref{kubo2WithConstantC}) to find
\begin{equation}
\frac{D_e}{D} =  \frac{\zeta_{\rm min}}{\langle \zeta\rangle_s}+ \frac{1}{\varepsilon}\frac{\ln 2}{\pi\langle \zeta\rangle_s}.
\end{equation}
and we thus recover within our complex formalism the result of Ref.~ \cite{man2017}. This result clearly agrees with the ergodic argument given earlier  when $\varepsilon\to \infty$. We see that for finite $\varepsilon$ the diffusion is faster for the reasons evoked above. However, this result is rather intriguing, firstly the fact that the correction scales as $1/\varepsilon$ is far from obvious. Another remarkable feature is that this correction depends only on the average height function through $1/\langle\zeta\rangle_s$ and moreover the associated prefactor $\ln(2)/\pi$ is universal. These analytical results are valid for any channel shape in the limit of wide channels, as will be confirmed in a  number of numerical tests in the following section.

\section{Numerical results}
\label{Sec7}
In this section we will show how the diffusion constant for periodic channels may be computed numerically. We use a standard finite element partial differential equation solver to (i) solve Laplace's equation Eq.~(\ref{lap}) with the integral constraint Eq.~(\ref{int}) and the boundary condition in Eq.~(\ref{bcs}) and (ii) carryout the subsequent evaluation of the effective diffusion constant via Eq.~(\ref{kubo1}). The results obtained are used to test some of the analytical results derived here and  elsewhere in the literature.

\subsection{The exact result of Zwanzig}

\begin{figure}
\begin{center}
  \includegraphics[scale=0.3]{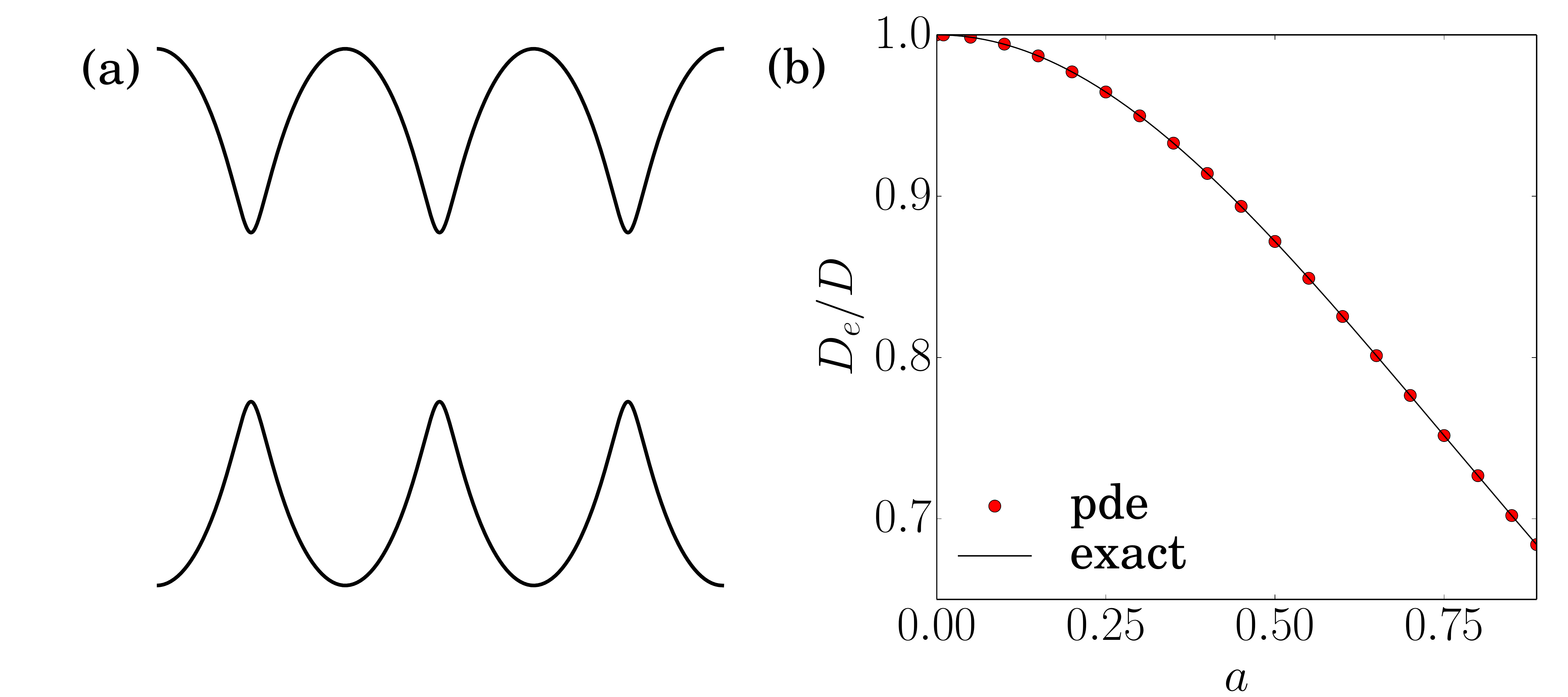}
  \caption{(color online) (a) Example of a channel in the class parameterized by Eqs.~(\ref{zx}) and (\ref{zy}), where $a= 0.5$ and $V=0.5$. (b) Comparison of the numerical evaluation of the system of Eqs.~(\ref{kubo1}-\ref{bcs}) (points) for the effective diffusion constant of the channel parameterized by Eqs.~(\ref{zx},\ref{zy}) and the exact result Eq.~(\ref{zex}) (solid line).
  }\label{zwanzigfig}
\end{center}
\end{figure}

In \cite{zwa1982} Zwanzig considered a symmetric channel parameterized using the curvilinear coordinate system defined by 
\begin{equation}
z=w+a \sin(w),
\end{equation}
which is equivalent to
\begin{eqnarray}
x&=& u + a \cosh(v)\sin(u) \label{zx} \\
y &=& v+ a \sinh(v)\cos(u),\label{zy}
\end{eqnarray} 
where the boundaries of the channel are given by $v=\pm V$. The height of the channel is single valued for $\alpha =a\cosh(V) <1$. An example of such a channel is shown in Fig.~\ref{zwanzigfig}a.

In this case Zwanzig showed that
\begin{equation}
\frac{D_e}{D} = \frac{1}{1 + \frac{\alpha^2\tanh(V)}{V}},\label{zex}
\end{equation}
which can be easily verified using Eq.~(\ref{exactzwanzig}). The existence of this exact result serves as a useful verification of the general formulas derived here for the diffusion equation (Eqs.~(\ref{kubo1}-\ref{bcs})) along with our numerical resolution of the associated partial differential equations. The effective diffusion constant is shown in Fig.~\ref{zwanzigfig}b for a channel with periodicity $2\pi$ with $V=0.5$ for 
values of $a$ taken between $0$ (the flat channel) through to the maximal value of $a$ for which the channel height is
a single valued function - the agreement is seen to be perfect. 

\subsection{Linear channel}

\begin{figure}
\begin{center}
  \includegraphics[scale=0.4]{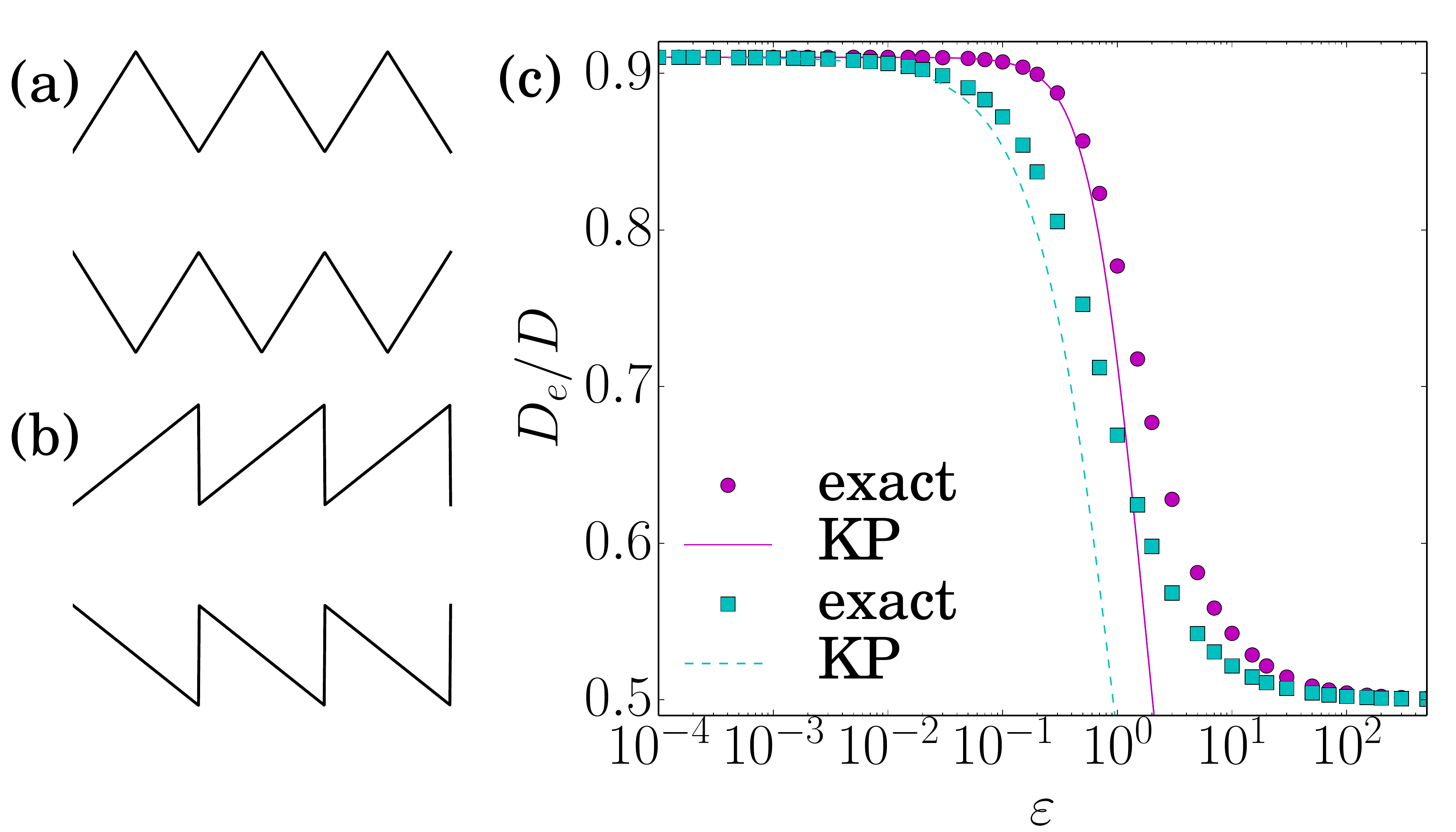}
  \caption{(color online) (a) Schematic of continuous symmetric piece wise linear channel (saw-tooth shape) (b) Schematic of discontinuous ratchet-like channel. 
  (c) Comparison of the (exact) numerical evaluation of the system of Eqs.~(\ref{kubo1}-\ref{bcs}) for the saw tooth channel for $a=0.25$ (circular symbols) and the ratchet channel for $a=0.5$ (square symbols) compared the Kalinay-Percus piece wise linear approximations given respectively by Eqs.~(\ref{lin1kp}) (solid line) and (\ref{lin2kp}) (dashed line).}\label{pwlfig}
\end{center}
\end{figure}

Here we consider dispersion in channels with piece wise linear profiles. Such channels are of interest as they are static/non-active realizations of the saw-tooth potentials often used to study ratchets and also because of the existence of the piece wise linear approximation scheme of Ref.~\cite{kal2006}, re derived here as Eq.~(\ref{pwl}). For concreteness we consider two linear channel profiles. We consider the periodic symmetric saw-tooth height profile $h(x) = \varepsilon\zeta(x)$ where $\zeta(x) = a +|x|$ for $x\in[-1/2,1/2]$ and shown in Fig.~\ref{pwlfig}a, already studied by \cite{ber2007,dag2010,ber2015,ver2016}, as well as the ratchet-like height profile $\zeta(x) = a + x$ on $[0,1]$ shown in Fig.~\ref{pwlfig}b. In the first case the channel height is continuous and Eq.~(\ref{pwl}) thus predicts that for the saw-tooth channel we have
\begin{equation}
\label{lin1kp}
\frac{D_e}{D} = \frac{2\tan^{-1}(\varepsilon)}{\varepsilon (4a+1) \ln(1+\frac{1}{2a})}
\end{equation}

However, for the ratchet channel the discontinuity in $\zeta$ means that Eq.~(\ref{pwl}) has some ambiguity. To solve the ambiguity we replace the discontinuity at $x=0,\ 1$ by defining 
a finite size $x_0$ and taking $\zeta$ to be a straight line between $(x_0, a+x_0)$ and $(1,a)$ (at the end of the calculation we take the limit $x_0\to 1$). This entails writing
\begin{equation}
\zeta(x)=
\begin{cases}
a+x ~~~&x\in[0,x_0] \\
a+\alpha(1-x) ~~~&x\in[x_0,1]
\end{cases},
\end{equation}
where $\alpha=\frac{x_0}{1-x_0}$ by continuity at $x_0$. Taking the limit $x_0 \to 1$ then recovers the ratchet channel. For a given value of $x_0$ we find
\begin{equation}
\begin{split}
\left\langle\frac{ \varepsilon\zeta'(x)}{\varepsilon\zeta(x)\tan^{-1}(\varepsilon \zeta'(x))}\right\rangle_s &= \frac{\varepsilon}{\tan^{-1}(\varepsilon)} \int_0^{x_0} \frac{dx}{a+x} + \frac{\alpha\varepsilon}{\tan^{-1}(\varepsilon \alpha)} \int_{x_0}^1 \frac{ dx}{\alpha(1-x)+a} \\
&= \frac{\varepsilon}{\tan^{-1}(\varepsilon)} \ln\left(1+\frac{x_0}{a}\right) + \frac{\varepsilon}{\tan^{-1}(\varepsilon \alpha)} \ln\left(1+\frac{x_0}{a}\right),
\end{split}
\end{equation} 
and taking the limit $x_0\to 1$ we find,
\begin{equation}
\left\langle \frac{ \varepsilon\zeta'(x)}{\varepsilon\zeta(x)\tan^{-1}(\varepsilon \zeta'(x))}\right\rangle_s=\varepsilon \left(\frac{2}{\pi} + \frac{1}{\tan^{-1}(\varepsilon)}\right) \ln\left(1+\frac{1}{a}\right),
\end{equation}
which yields
\begin{equation}
\label{lin2kp}
\frac{D_e}{D} = \frac{1}{\varepsilon \left(a+\frac{1}{2}\right) \left(\frac{2}{\pi} + \frac{1}{\tan^{-1}(\varepsilon)}\right) \ln\left(1+\frac{1}{a}\right)}.
\end{equation}

The comparison of the piece wise linear approximation with the exact results is shown in Fig.~\ref{pwlfig}c we see that  the approximation clearly fails for the ratchet channel, since the deviations from the basic FJ approximation are not well captured by the piece wise linear approximation. In turn, the  piece wise linear approximation is reasonable for the saw-tooth profile for values of $\varepsilon$ of order $1$. This may linked to the fact that this approximation is exact at order $\varepsilon^2$. In both cases, the piece wise linear approximation underestimates the effective diffusion constant for large $\varepsilon$. Indeed, in general the piece wise approximation predicts that the effective diffusion constant behaves as $D_e \sim 1/\varepsilon$ for large $\varepsilon$. This is at variance with the exact numerical calculation and the large $\varepsilon$ analysis of section \ref{largee} where the effective diffusion constant saturates to a non-zero asymptotic limit ($D_e \to a/(a+1/4)$ for the saw-tooth and $D_e\to a/(a+1/2)$ for the ratchet). For the parameters $a=1/4$ for the saw tooth case and $a=1/2$ for the ratchet, the predicted large $\varepsilon$ result is $D_e\to 1/2$ for both cases, in perfect agreement with the numerical results shown in Fig.~\ref{pwlfig}.

\subsection{Smooth channel}

\begin{figure}
\begin{center}
  \includegraphics[scale=0.2]{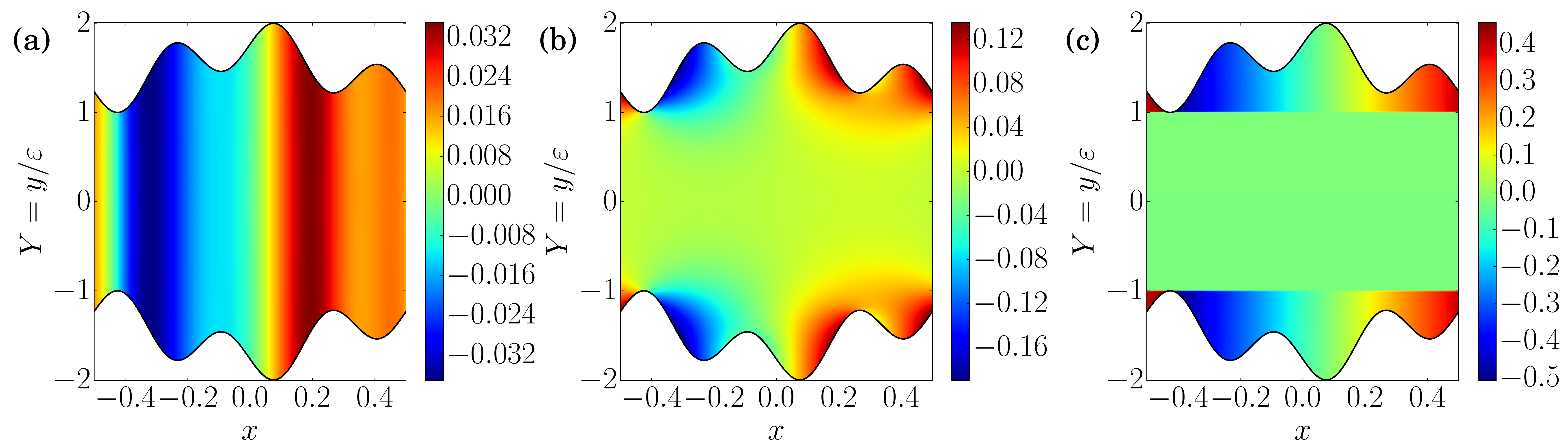}
  \caption{(color online) (a) Numerical solution of $f(x,Y)$ for $\varepsilon=0.01$. The function is independent of the coordinate $Y=y/\varepsilon$ in accordance with the FJ approximation. The resulting diffusion constant is thus well approximated by $D_e \sim D_{\rm FJ}$ given in Eq.~(\ref{fj}). (b) Numerical solution of $f(x,Y)$ pour $\varepsilon=0.5$. The function is no longer independent of $Y$ and is also non-zero in the region of the canal $|Y|<\zeta_{\rm min}$. (c) Numerical solution of $f(x,Y)$ for $\varepsilon=100$. In agreement with the results of section \ref{largee} we see that $f=0$ in the region of the channel $|Y|\le \zeta_{\rm min}$ and is independent of $Y$ pour $|Y|>\zeta_{\rm min}$}.\label{sinefigs}
\end{center}
\end{figure}

In this section we consider the case of a smoothly varying channel where all terms appearing in the perturbation series
solution for the effective diffusion constant $D_e$, derived in section \ref{pert} are finite. In particularly we will show how a formula, which fits the numerical evaluation for the diffusion constant, can be derived for all values of $\varepsilon$ using a Pad\'e type re summation of the small $\varepsilon$ perturbation expansion along with the first two terms (the ${\cal O}(1)$ and ${\cal O}(1/\varepsilon)$ ones) in 
the large $\varepsilon$ expansion derived in section \ref{largee}. In addition we will demonstrate numerically certain of the key results derived in this paper in the large and small $\varepsilon$ regimes. 

For the purposes of this section we consider the channel defined by
\begin{equation}
\zeta(x)=1.5+0.266\left[\cos(2\pi x)+\sin(6\pi x)\right].\label{esch}
\end{equation}
In order to compare numerical results for different values of $\varepsilon$ we define $Y = y/\varepsilon$. Shown in Figs.~\ref{sinefigs}a-c is the numerically evaluated function $f(x,Y)$ for values of $\varepsilon$ given respectively by $\varepsilon= 0.01,\ 0.5$ and $100$. First, at small values of $\varepsilon$ we see that the solution $f(x,Y)$ shown in 
Fig.~\ref{sinefigs}a is independent of $Y$, this is the underlying hypothesis behind the FJ approximation and we see that for this narrow channel the approximation is works well. On increasing $\varepsilon$ to $0.5$, in Fig.~\ref{sinefigs}b, we see that the solution develops a dependence on $Y$. Finally for large $\varepsilon$, Fig.~\ref{sinefigs}c, we see that the analysis of section \ref{largee} is confirmed. Firstly we see that the solution vanishes within the part of the channel $|Y|<\zeta_m$ to high numerical accuracy, while the solution becomes, as for the small $\varepsilon$ case, independent of $Y$ in the region $|Y|>\zeta_m$ and takes the approximate form $f(x,y) \approx x$.

We now consider the accuracy of the perturbation theory developed in this paper based on 
the small $\varepsilon$ expansion, which for the smooth channel considered here is always
well defined. Shown in Fig.~\ref{sinediff}a is the result of the perturbation theory starting at order
${\cal O}(\varepsilon^2)$, through to that at ${\cal O}(\varepsilon^6)$ along with the exact numerical evaluation of the diffusion constant. The results clearly improve with increasing the order of the perturbation
expansion -- however, they all depart strongly from the exact result when $\varepsilon$ is ${\cal O}(1)$. Furthermore the perturbation expansion appears to give lower bounds when 
taken at ${\cal O}(\varepsilon^2)$ and ${\cal O}(\varepsilon^6)$ and upper bounds when taken at 
${\cal O}(\varepsilon^4)$ (without any re summation). This alternating structure in the perturbation expansion is also seen for diffusion in random Gaussian potentials \cite{dea94, dea95}. The problem with this alternating structure of the perturbation theory is that the diffusion constant will either become negative or greater than $D$ (which is clearly not possible on physical grounds) depending on the order of perturbation theory employed. In addition, we have seen that $D_e$ saturates at large values of $\varepsilon$ 
which is incompatible with a pure polynomial expansion in $\varepsilon$. Based on 
the ${\cal O}(\varepsilon^2)$ result of perturbation theory a number of authors have proposed 
the following alternative forms, which of course agree at ${\cal O}(\varepsilon^2)$ with the basic perturbation theory result. In our notation these alternative expressions are 
\begin{equation}
\frac{D_e}{D} = \frac{1}{\langle \zeta \rangle_s \langle \zeta^{-1} \rangle_s} \frac{1}{1 + \varepsilon^2 \frac{\langle \zeta'^2/\zeta\rangle_s}{3 \langle \zeta^{-1} \rangle_s}},\label{rszw}
\end{equation}
proposed in \cite{zwa1991}. Another alternative is given by
\begin{equation}
\frac{D_e}{D} = \frac{1}{\langle \zeta \rangle_s}\langle \frac{(1+\varepsilon^2\zeta'^2)^{\frac{1}{3}}}{\zeta}\rangle_s^{-1},\label{rsrr}
\end{equation}
which was given in \cite{reg2001}. A third alternative is given by Eq.~(\ref{pwl}), originally proposed in \cite{kal2006}. The comparison of these approximations with the exact numerical evaluation for the diffusion constant for the case in hand is shown in Fig.~\ref{sinediff}b. The advantage of these formulations is that the effective diffusion constant 
is now bounded within the region $[0,D]$. However, in the limit of large $\varepsilon$ all these
proposed forms predict that the diffusion constant becomes zero and all are significantly different from the exact result for $\varepsilon$ of order 1. The last question we will therefore address here is whether there is a way to tie up the small and large $\varepsilon$ results derived here to obtain a formula describing the effective diffusion constant for all
values of $\varepsilon$? Below we will pursue a Pad\'e like approximant scheme which
exploits both regimes of the perturbation theory studied in this paper.

We begin by writing the small $\varepsilon$ expansion for the effective diffusivity given by Eq.~(\ref{o4}) 
\begin{equation}
\begin{split}
\frac{D_e}{D} &= \frac{1}{\langle \zeta \rangle_s \langle \zeta^{-1} \rangle_s} \left[1-\frac{\langle \zeta'^2/\zeta \rangle_s}{3\langle \zeta^{-1} \rangle_s} \varepsilon^2 + \left(\frac{\langle \zeta'^2/\zeta \rangle_s^2}{9\langle \zeta^{-1} \rangle_s^2} + \frac{4 \langle \zeta'^4/\zeta \rangle_s + \langle \zeta''^2 \zeta \rangle_s}{45 \langle \zeta^{-1} \rangle_s}\right) \varepsilon^4 + ...\right] \\
&=D_{\rm FJ} \left[1+\lambda_2 \varepsilon^2 + \lambda_4 \varepsilon^4 + ...\right],
\end{split}
\end{equation}
where $\lambda_i =\frac{C_i}{C_0}$. In principle one can gain in precision by also using the ${\cal O}(\varepsilon^6)$ correction, though the resulting algebraic complication does
not really justify the gain in accuracy.
Using the large $\varepsilon$ expansion developed in section \ref{largee} we also write
\begin{equation}
\frac{D_e}{D} = \frac{1}{\langle \zeta \rangle_s} \left[\zeta_{\rm min}+ \frac{\ln 2}{\pi \varepsilon}+ ...\right] = D_{\infty} \left[1+ \frac{\alpha}{\varepsilon}+ ...\right].
\end{equation}
A Pad\'e type re summation is constructed by writing
\begin{equation}
\frac{D_e}{D} = D_{\rm FJ} \frac{1+ a_1 \varepsilon + a_2 \varepsilon^2 +a_3 \varepsilon^3}{1+ b_1 \varepsilon + b_2 \varepsilon^2 +b_3 \varepsilon^3},
\end{equation}
where the $a_i$ and $b_i$ are obtained from the two expansions, for small and large $\varepsilon$. The coefficients are found to be given by
\begin{equation}
\begin{split}
a_1&=b_1=\frac{[\lambda_2^2+\lambda_4(K-1)](K-1)}{\lambda_2^2 K \alpha} \\
a_2&=\lambda_2 - \frac{\lambda_4}{\lambda_2} \\
b_2&=- \frac{\lambda_4}{\lambda_2} \\
a_3&=Kb_3 = \frac{\lambda_2^2+\lambda_4(K-1)}{\lambda_2 \alpha} \\
K&=\frac{D_{\infty}}{D_{\rm FJ}}.\label{pade}
\end{split}
\end{equation}

The results of this approximation for the class of channels described by Eq.~(\ref{esch}) is shown in Fig.~\ref{sinediff}b. We see that the resulting approximation is in very good agreement with the exact numerical results and that, as is to be expected, the maximum
errors arise in the crossover region where $\varepsilon$ is of order $1$.
\begin{figure}
\begin{center}
  \includegraphics[scale=0.3]{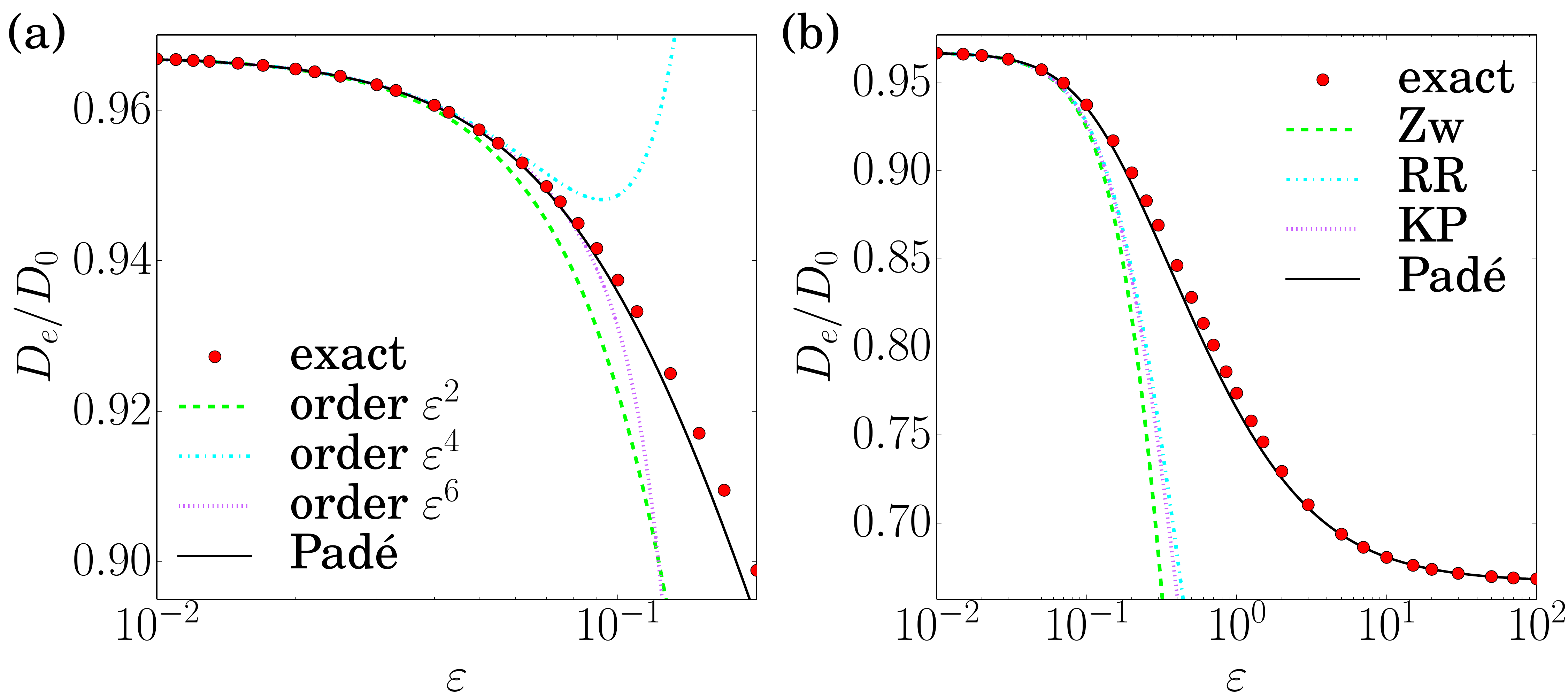}
  \caption{(color online) (a) Exact numerical value of $D_e$ (circles) for the class of channels described 
  by Eq.~(\ref{esch}) as a function of $\varepsilon$ compared with perturbation theory 
  in $\varepsilon^2$ at order $1$ (dot-dashed line), $2$ (dashed line) and $3$ (solid line).
  (b) Same exact numerical values for $D_e$ as in (a) (circles) compared with the ${\cal O}(\varepsilon^2)$ re summations given by Eq.~(\ref{rszw}) due to Zwanzig \cite{zwa1991} Eq.~(\ref{rsrr}) (dashed line), due to Regura and Rubi \cite{reg2001} (dot-dashed line) and Eq.~(\ref{pwl}) due to Kalinay and Percus \cite{kal2006}
  (dotted line). Shown as a solid line is the Pad\'e approximant scheme proposed here
  in the system of Eqs.~(\ref{pade})
 }\label{sinediff}
\end{center}
\end{figure}

\section{Conclusion}
In this paper we have examined the effective dispersion of Brownian particles in a channel geometry. Rather than attempting to write an effective one-dimensional diffusion equation to describe the system as in many past studies, we have exploited a direct formula for the effective diffusion constant which is obtained from the system of time independent equations Eqs.~(\ref{kubo1}-\ref{bcs}) which were derived here for completeness but are in agreement with a general theory of dispersion coefficients developed in \cite{gue2015a} and \cite{gue2015b} and also with the macro transport theory of Brenner \cite{bre1993} when applied to the channel problem \cite{dorfman2014assessing}. Using a formulation in terms of analytic functions [Eqs.~(\ref{kubo1},\ref{lap},\ref{int},\ref{bcs})], we were able to develop a compact formulation of the perturbation theory in channel width scale $\varepsilon$. Notably we were able to show that the homogenization scheme proposed by Kalinay and Percus \cite{kal2006} gives the correct diffusion constant at order $\varepsilon^6$. The exact results of Zwanzig for parameterized channels and the piece wise linear approximation of \cite{kal2006} can 
be deduced from the formalism developed here in a very compact manner. Most studies on diffusion in channels are focused on narrow channels. The formalism developed here was also exploited to study wide channels, yielding a physically intuitive result in the limit or very large channels but also giving the first non-trivial correction which takes a remarkably universal form. Finally we were able to show how the small and large $\varepsilon$ expansions for the effective diffusion constant can be combined in a Pad\'e type approximant which accurately describe continuous channels for all values of $\varepsilon$. 

There are a number of potential extensions of this work based on the static approach proposed in \cite{gue2015a,gue2015b} which can be extended to systems having a current in their equilibrium states. This means that the approach could be applied to dispersion in channels with driving forces both along the channel and normal to the channel \cite{gho2010,bur2008,burada2009diffusion}(arising from gravity for example). Furthermore the approach of \cite{gue2015a,gue2015b} can also be adapted to extract the finite time behavior of transport coefficients and could be applied to the problem of diffusion in channels.

\appendix
\section{Derivation of Eq.~(\ref{kubo2WithConstantC})}
\label{AppendixConstantC}

The function $w(x+iy)$ is analytic, thus for $z=x+iy$, we have for all closed paths $\mathcal{C}$
\begin{align}
\int_\mathcal{C} dz w(z) =0.
\end{align}
Note that here we can add an arbitrary constant to $w(z)$ and of course the above remains trivially true (justifying making the choice $v_0=0$ in Eq.~(\ref{vo})). If we choose $\mathcal{C}$ to be the boundary of one periodic cell, with $x \in [0,L]$, we get
\begin{align}
\int_\mathcal{C} dz w(z) &= \int_0^L [dx + i h'(x) dx] [u(x,h(x))+iv(x,h(x))] + \int_{h(L)}^{-h(L)} idy [u(L,y) + i v(L,y)] \nonumber \\
&+ \int_L^0 [dx - i h'(x) dx] [u(x,-h(x))+iv(x,-h(x))] + \int_{-h(0)}^{h(0)} idy [u(0,y) + i v(0,y)]
\end{align}
If we use the symmetries $u(x,-y) = u(x,y)$ and $v(x,-y) = -v(x,y)$, and the periodicity in $x$ of all functions, the expression becomes
\begin{align}
\int_\mathcal{C} dz w(z)= 2 i \int_0^L dx [h'(x) u(x,h(x)) + v(x,h(x))]=0.
\end{align}
From  Eq.~(\ref{kubo2}), we obtain
\begin{align}
\frac{D_e}{D}= 1+ \frac{2}{\Omega} \int_0^L dx h'(x) u(x,h(x)) = 1- \frac{2}{\Omega} \int_0^L dx v(x,h(x)),
\end{align}
and the boundary condition (\ref{be}) yields
\begin{align}
\frac{D_e}{D}= 1- \frac{2}{\Omega} \int_0^L dx [h(x) - C] = \frac{C}{\langle h \rangle_s}.
\end{align}

\end{document}